\documentclass[reprint,twocolumn,prl,showpacs, amsmath, amssymb, superscriptaddress,aps]{revtex4-2}

\usepackage{graphicx}% Include figure files
\usepackage{dcolumn}% Align table columns on decimal point
\usepackage{bm}% bold math
\usepackage{bbold}
\usepackage{amsmath}
\usepackage{comment}
\usepackage{xcolor}

\newcommand{\id}{\mathbb{1}} % unit matrix symbol

\begin{document}

\preprint{APS/123-QED}

\title{Exceptional points of PT-symmetric reflectionless states in complex scattering systems}
\author{Cl{\'e}ment Ferise}
\affiliation{Univ Rennes, CNRS, IETR - UMR 6164, F-35000 Rennes, France}%

\author{Philipp del Hougne}
\affiliation{Univ Rennes, CNRS, IETR - UMR 6164, F-35000 Rennes, France}%

\author{Simon F{\'e}lix}
\affiliation{
Laboratoire d'Acoustique de l'Universit{\'e} du Mans (LAUM), UMR 6613, Institut d'Acoustique - Graduate School (IA-GS), CNRS, Le Mans Universit{\'e}, France
}%

\author{Vincent Pagneux}
\affiliation{
Laboratoire d'Acoustique de l'Universit{\'e} du Mans (LAUM), UMR 6613, Institut d'Acoustique - Graduate School (IA-GS), CNRS, Le Mans Universit{\'e}, France
}%

\author{Matthieu Davy}
\affiliation{Univ Rennes, CNRS, IETR - UMR 6164, F-35000 Rennes, France}%
%\email{matthieu.davy@univ-rennes1.fr}
\date{\today}

\begin{abstract} %Maximum of 150 words, no references
We investigate experimentally and analytically the coalescence of reflectionless (RL) states in symmetric complex wave-scattering systems. We observe RL-exceptional points (EPs), first, with a conventional Fabry-Perot system for which the scattering strength within the system is tuned symmetrically, and then with single- and multi-channel symmetric disordered systems. We identify that an EP of the parity-time (PT)-symmetric RL-operator is obtained when the spacing between central frequencies of two natural resonances of the system is equal to the decay rate into incoming and outgoing channels. Finally, we leverage the transfer functions associated with RL and RL-EPs states to implement first- and second-order analog differentiation.

%We investigate analytically and experimentally exceptional points (EPs) associated with parity-time (PT) reflectionless (RL) states in symmetric complex wave-scattering systems. We find that eigenvalues and eigenstates of RL-states coalesce when the spacing between central frequencies of two natural resonances of the system is perfectly balanced by the decay rate of incoming and outgoing channels. We observe in the microwave regime RL-exceptional points, first, with a conventional Fabry-Perot system for which the scattering strength is tuned symmetrically, and then with single- and multi-channel symmetric disordered systems. Finally, we leverage the transfer functions associated with RL and RL-EPs states to implement first- and second-order analog differentiation.

\end{abstract}

%\maketitle

\begin{titlepage}
  \maketitle
\end{titlepage}

Exceptional points (EP) are spectral singularities in non-Hermitian systems at which two or more eigenvalues and eigenstates coalesce \cite{Heiss2000,Berry2004,RotterI2009,Moiseyev2011,Cao2015}. EPs have mainly been explored for resonant states that are the poles of the scattering matrix and scattering states. In systems with losses only, finding an EP between two resonant modes requires that the mutual coupling between resonances satisfy a critical relation with their loss factors \cite{Miri2019}. EPs are however more easily realizable in systems with PT-symmetry obtained by balancing gain and loss in symmetric regions \cite{Ozdemir2019,Miri2019}. Many unconventional features of EPs have been demonstrated theoretically and experimentally such as the design of unidirectional invisibility  \cite{Lin2011,Fleury2015}, asymmetric mode switching \cite{Doppler2016} or directional lasing \cite{Peng2016}, to cite a few (see Refs.~\cite{El-Ganainy2018,Miri2019} for reviews). These spectral singularities are also interesting for sensing applications since the energy (or frequency) splitting between $n$ degenerate eigenstates scales as the $n$th root of the perturbation \cite{Chen2017,Hodaei2017}.   

Recently, a new kind of EPs associated with reflectionless (RL) scattering states rather than resonances has been investigated. RL states are eigenstates of a non-Hermitian operator $H_{\mathrm{RL}}$ based on the wave equation with incoming channels connected to a scattering system modeled as gain and outgoing channels modeled as losses~\cite{Dhia2018,Sweeney2020,Stone2020}. The eigenvalues $\Tilde{\omega}_R$ of $H_{\mathrm{RL}}$ are distinct from the resonance spectra related to the poles of the scattering matrix $S(\omega)$. When an eigenvalue is tuned to the real axis, the corresponding RL-state enable reflectionless coupling of incoming channels. The excitation of multi-channel RL-states in disordered matter requires non-trivial wavefront shaping~\cite{Pichler2019}.

A special case of RL-EPs found for purely incoming boundary conditions is the coalescence on the real axis of two zeros of the complete scattering matrix $S(\omega)$ known as a perfectly absorbing EP~\cite{Sweeney2019,Suwunnarat2021,wang2021}. The phenomenon of coherent perfect absorption (CPA) occurs when absorption within a scattering medium balances the excitation rate of incoming channels~\cite{Li2017,Fyodorov2017,Baranov2017,Krasnok2019}. CPA is the time-reverse of lasing at threshold~\cite{Chong2010,Roger2015} and a generalization of the critical coupling condition \cite{Piper2014,Baranov2017}. CPA has been implemented in a wide range of regular scattering systems~\cite{wan2011time}, as well as in disordered matter~\cite{f2020perfect,chen2020perfect,frazier2020wavefront,delHougne2020CPA,del2021coherent,Chen2021}. The latter has been achieved in particular through purposeful perturbations of the scattering system, which can include additional constraints, such as on frequency or CPA wavefront~\cite{f2020perfect,frazier2020wavefront,delHougne2020CPA,del2021coherent}. When two CPA-states coalesce, the broadened lineshape of the absorption spectrum indicates the existence of perfectly absorbing EPs \cite{Sweeney2019,Suwunnarat2021,wang2021}.

In flux-conserving systems, zero reflection and therefore perfect transmission of an incoming wavefront occurs when the decay rates into incoming and outgoing channels are equal. The probability of finding RL scattering modes is naturally enhanced in scattering systems with mirror-symmetry for which the RL-operator is PT-symmetric and RL-eigenvalues feature an exciting property. An RL-eigenvalue remains on the real axis under continuous perturbation of the system until it coalesces with a second one at an EP before splitting into a complex-conjugate pair \cite{Dhia2018,Sweeney2019,Stone2020}. %A signature of a RL-EP on the real axis is the  and reflection zero. 
The spectral broadening of the transmission peak corresponding to RL-EPs has been observed in multimirror cavities \cite{VandeStadt1985,Stone1990,Stephen2017} and in numerical simulations of quantum dots \cite{Joe2000} and atomic wires \cite{Lee2001}, even though not interpreted as an EP. However, a clear experimental demonstration of RL-EPs in disordered systems and an analysis of RL-states in term of natural resonances has to date not been reported.

\begin{figure*}
    \centering
    \includegraphics[width=18cm]{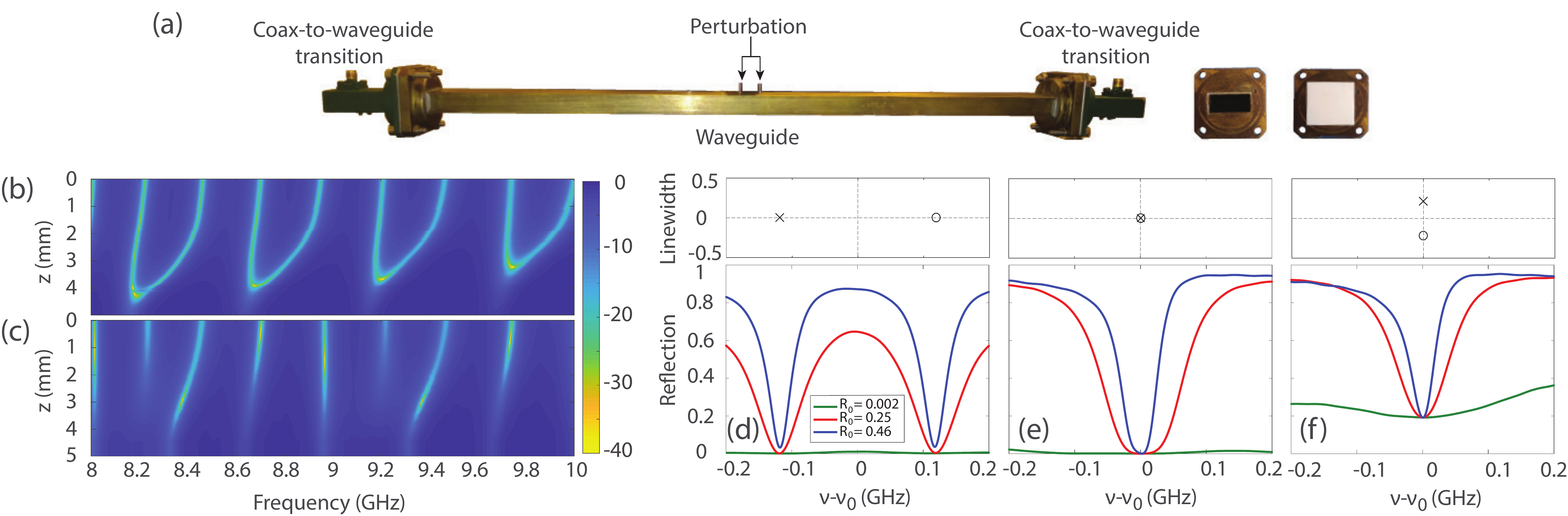}
    \caption{(a) Photography of the experimental setup which is a rectangular waveguide supporting only the fundamental transverse-electric mode between 7 and 11 GHz. The penetration depth $z$ of two rods inserted symmetrically near the middle of the waveguide can be finely controlled with a translation stage. Inset: a channel (coax-to-waveguide transition) with an empty opening or with an alumina slab placed in front of it to increase the reflectivity at the interfaces of the waveguide. (b-c) Colorscale representation of the reflection on a dB-scale, $10 \mathrm{log}[R(\nu,z)]$, measured with a vector network analyzer between 8 and 10 GHz for a symmetric (b) and a asymmetric (c) perturbation. RL-EPs appear only in the symmetric case when two branches with $R\sim0$ shown in yellow coincide. (d-f) $R(\nu,z)$  in the underperturbed $z<z_{\mathrm{EP}}$ (d), critically perturbed $z = z_{\mathrm{EP}}$ (e) and overperturbed  $z>z_{\mathrm{EP}}$ (f) regimes as a function of rescaled frequency $\nu - \nu_0$ ($\nu_0 = 8.2$~GHz). The spectra are shown for zero (green line), one (red line) and two (blue line) dielectric slabs at each interface. The corresponding RL-eigenvalues found from Eq.~(\ref{eq:eigenvalues}) are shown on the top in the complex plane.}
    \label{fig:setup}
\end{figure*}

In this article, we  observe the existence of RL-EPs in symmetric regular and disordered systems that are tuned toward this scattering anomaly by inserting a symmetric defect. We first start with a multimirror Fabry-Perot cavity for which the reflectivity at the center is progressively increased. Our harmonic analysis reveals that an RL-EP is obtained when the spacing between the central frequencies of two quasi-normal modes (QNMs) is equal to their linewidth. Our experimental results in the microwave range are in excellent agreement with a two-level model. We then demonstrate RL-EPs in disordered single- and multi-channel waveguides. Finally, we show that designing systems operating at an RL-EP is relevant to analog computations of derivatives.

We measure spectra of the reflection coefficient $r(\nu)$ and the transmission coefficient $t(\nu)$ through a single-mode rectangular waveguide (length $L=400$~mm, width $W=22.86$~mm and height $H=10.16$~mm) with two coax-to-waveguide transitions attached to the openings (see Fig.~\ref{fig:setup}(a)). A dielectric alumina slab of reflectivity $R_0 = 0.25$ (see SM) is positioned between each transition and the waveguide to increase the internal reflection at the interfaces. A perturbation is introduced symmetrically with respect to the center of the waveguide by inserting two aluminium rods (diameter $2$~mm) through two holes drilled into the top plate and spaced by $12$~mm. The penetration depth $z$ varies from 0 to $8$~mm in steps of $\Delta z = 0.02$~mm ($\Delta z \sim \lambda / 1666$). This system is therefore equivalent to a multi-mirror Fabry-Perot interferometer with tunable reflectivity at the center. 

A colormap of the reflection $R(\nu,z) = |r(\nu,z)|^2$ is shown in Fig.~\ref{fig:setup}(b) on a log-scale. In absence of the perturbation ($z = 0$), the frequencies corresponding to extremely small reflection (bright area) are regularly spaced as expected for a Fabry-Perot interferometer. As the penetration depth $z$ increases, the RL-frequencies move closer until they coalesce for a critical perturbation ($z = z_{\mathrm{EP}}$) to form RL-EPs. The two peaks on the spectrum of $R(\nu)$ found in the underperturbed regime ($z<z_{\mathrm{EP}}$) transform into a single broadband one at $z_{\mathrm{EP}}$ with a flattened quartic line shape which is characteristic of EPs \cite{Sweeney2019,Suwunnarat2021,sol2021meta,wang2021}. Once two RL-eigenvalues have collapsed, they leave the real axis as complex-conjugate pairs. In the overperturbed regime $z>z_{\mathrm{EP}}$, the minimum of $R(\nu)$ increases with $z$ as the imaginary part of RL-eigenvalues moves away from the real axis. In contrast, for an asymmetrical perturbation of the Fabry-Perot cavity (a single rod inserted at $x = L/4$), the zero-reflection frequencies do not collapse but move independently in the complex plane (see Fig.~\ref{fig:setup}(c)). 

The same procedure is then repeated for samples with zero and two alumina slabs at each interface, yielding $R_0 = 0.002$ and $R_0 = 0.46$. For the smallest $R_0$, the reflectivity is solely due to the small impedance mismatch between the transitions and the waveguide. In each case, a symmetric perturbation leads to the formation of RL-EPs. The linewidth of resonances decreases with increasing $R_0$ and the spectral dips corresponding to RL-states narrow as seen in Fig.~\ref{fig:setup}(d-f).

\begin{figure}
    \centering
    \includegraphics[width=8.5cm]{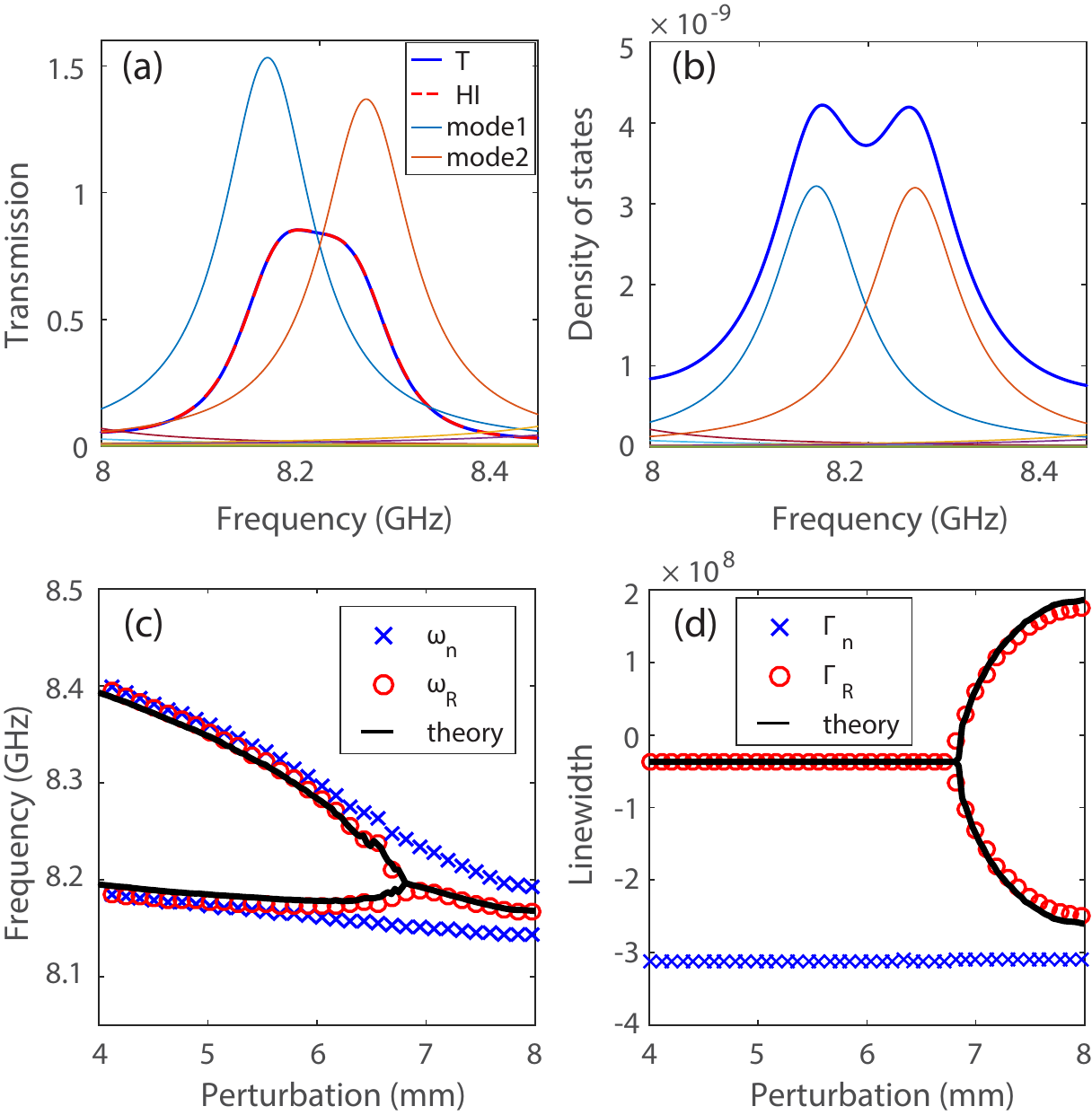}
    \caption{(a) Modal analysis of the transmission coefficient $t(\nu,z_{\text{EP}})$. Two modes with lines shown in orange and cyan mainly contribute to $T = |t(\nu,z_{\text{EP}})|^2$ (blue line) which is perfectly reconstructed from harmonic inversion (red dashed line). (b) the DOS present two peaks corresponding to the maxima of the two resonances. (c,d) Real (c) and Imaginary (d) parts of the reflectionless eigenvalues (red circles) and eigenfrequencies (blue crosses) with respect to $z$. The black lines are the theoretical values of $\Tilde {\omega}_R$ found from Eq.~(\ref{eq:eigenvalues}) in which the values of $\omega_n(z)$ and $\gamma(z)$ are injected.
    }
    \label{fig:modes}
\end{figure}

We now analyze RL-eigenvalues found by tracking the local minima of $R(\omega,z)$ in terms of natural resonances of the medium. The QNMs are solutions of the wave equation with purely outgoing boundary conditions (i.e. for \textit{both} left and right channels). The eigenstates $\psi_n(x)$ are associated with complex frequencies $\Tilde{\omega}_{n} =\omega_n -i\Gamma_n/2 $ with central frequency $\omega_n$ and linewidth $\Gamma_n$. We extract the set of complex frequencies $\Tilde{\omega}_{n}$ from a modal analysis of transmission spectra between 7 and 11 GHz using the harmonic inversion method \cite{Kuhl2008,Davy2018} (see SM). In absence of the perturbation, the central frequencies almost coincide with RL-frequencies. However, for $z = z_{EP}$, two overlapping resonances mainly contribute to the flattened transmission peak as seen in Fig.~\ref{fig:modes}(a). Two peaks are observed on the spectrum of the density of states (DOS) $\rho = (1/2\pi) \Sigma_n (\Gamma_n/2) / [(\omega-\omega_n)^2+(\Gamma_n/2)^2]$ in Fig.~\ref{fig:modes}(b) with a spacing between central frequencies approximately given by the linewidths. Even for the highest penetration depth of the rods, the resonances do not coalesce.

We now derive analytical expressions of $\Tilde{\omega}_{n}$ and RL-eigenvalues $\Tilde{\omega}_{\mathrm{R}}$ using a two-level model. The effective Hamiltonian $H_{\mathrm{eff}}$ of the open system is expressed in the basis of two successive modes of a closed one-dimensional waveguide, $H_{\text{eff}} = H_0 -i V_0 V_0^T/2 -iV_1 V_1^T /2$, with $H_0$ being a diagonal matrix with elements $\omega_0 - \delta\omega_0/2$ and $\omega_0 + \delta\omega_0/2$. The left and right vectors $V_0$ and $V_1$ account for the coupling of the closed system to the channels and are real for systems with time-reversal symmetry. The scattering matrix $S(\omega)$ writes $S(\omega) = 1-iV^T [\omega \id - H_{\text{eff}}]^{-1} V$, where $V = [V_0 \; V_1]$. For an empty one-dimensional cavity of length $L$, the $n$th eigenfunction of the closed systems is $\psi_n(x) = (2/L) \mathrm{sin}(n\pi x/L)$. We consider here an even resonance ($n \equiv 0 \;(\bmod\; 2)$) and an odd resonance ($n \equiv 1 \;(\bmod\; 2)$). The coupling vectors are related to the derivative of $\psi_n(x)$ at $x = 0$  and $x = L$. $V_0$ is therefore symmetric, $V_0 = \sqrt{\gamma/2}(1~1)$ and $V_1$ is anti-symmetric $V_1 = \sqrt{\gamma/2}(1~-1)$. The coupling rate $\gamma/2$ of each channel to the cavity depends on the reflectivity at the interfaces of the waveguide.  The effective Hamiltonian is therefore a diagonal matrix  with eigenvalues $\Tilde {\omega}_{M\pm} = \omega_0 \pm \delta\omega_0/2 - i\gamma/2$ (see SM).

Inserting a rod in the middle of the waveguide leads to a local change $\Delta \epsilon (x=L/2)$ of the permittivity. For a small penetration depth, the frequency shift of the two resonances can be calculated using standard perturbation theory. At first order, the shift is $\Delta \omega_n \propto -\Delta \epsilon(x) |\psi_n(x)|^2 $ \cite{Cotrufo,Cogne2019}. For an even resonance, the field cancels at the center of the waveguide and $\Delta \omega_n \sim 0$. For odd resonances, $|\psi_n(x)|^2\neq 0$ and the central frequency shifts towards smaller values. As the penetration depth increases, resonances come together in pairs with central frequencies and spacings that are functions of $z$, $\omega_0(z)$ and $\delta\omega_0(z)$. 

RL-modes are eigenvalues of the PT-symmetric operator $H_{\mathrm{R}} = H_0 + i V_0 V_0^T /2 -i V_1 V_1^T/2$, where an effective gain is associated to incoming channels at the left interface. The coupling of the channels now results in anti-diagonal terms $i\gamma$: 
\begin{equation}
    H_{\mathrm{R}} = \omega_0(z) \id + \frac{1}{2}\begin{pmatrix}  - \delta \omega_0(z)  &  i\gamma\\  i\gamma &  \delta \omega_0(z)   \end{pmatrix} 
\label{eq:HR}
\end{equation}
The RL-eigenvalues $\Tilde {\omega}_R$ are then found from a diagonalization of $H_{\mathrm{RL}}$:
\begin{equation}
\Tilde {\omega}_{R\pm} = \omega_0(z) \pm \frac{1}{2}\sqrt{\delta\omega_0(z)^2 -\gamma^2}.
\label{eq:eigenvalues}
\end{equation}
Uniform absorption within the waveguide is incorporated by adding an imaginary part $-i\gamma_a$ to complex frequencies. We estimate from the modal analysis that $\gamma_a = 30$~MHz (see SM).

When the coupling of channels is small compared to the spacing between resonances, $\gamma \ll \delta\omega_0$, RL-eigenvalues are real and coincide with the central frequencies of resonances $\omega_n$. However, the spacing between RL-eigenvalues decreases more rapidly than $\delta\omega_0(z)$ as the perturbation strength is symmetrically tuned. A reflectionless EP is found when losses through channels are equal to the spacing between the two resonances, $\delta\omega_0(z_{\text{EP}}) = \gamma$. For larger perturbations, the two RL-eigenvalues become a complex-conjugate pair. Assuming that $\delta\omega_0(z)$ scales linearly with the penetration depth $z$ near the EP, $\delta\omega_0(z) = \gamma+\kappa/2(z_{\text{EP}}-z) $  (see SM), gives a splitting between RL-eigenvalues for $z < z_{\text{EP}}$ near the EP $\Tilde {\omega}_{R+} - \Tilde {\omega}_{R-} \sim \sqrt{\gamma \kappa (z_{\text{EP}}-z)} $ which is characteristic of the square-root detuning behavior of EPs under small perturbations. The same splitting is found on imaginary parts of $\Tilde {\omega}_{R}$ for $z > z_{EP}$.

The experimental results in Fig.~\ref{fig:modes}(c) for $\text{Re}[\Tilde {\omega}_R]$ are in excellent agreement with the prediction of Eq.~(\ref{eq:eigenvalues}) in which we inject the values of $\omega_0(z)$, $\gamma(z)$ and $\delta\omega_0(z)$ extracted from the modal analysis. This agreement also highlights the effectiveness of the two-level model. For $z<z_{\mathrm{EP}}$, the imaginary part of $\Tilde {\omega}_R$ is equal to the absorption decay rate $\Gamma_{R} = - \gamma_a$. For $z>z_{\mathrm{EP}}$, we fit $R(\omega = \omega_0)$ using the following analytical expression for the reflection coefficient  %$r = 1-iV_0^T [\omega \id - H_{\text{eff}}]^{-1} V_0$ 
(see SM for derivation):
\begin{equation}    \label{eq:r}
\begin{split}
   r & = \frac{[\omega-\omega_0(z)]^2+[\gamma^2-\delta\omega_0(z)^2]/4}{[\omega-\omega_0(z)]^2-[\delta\omega_0(z)^2+\gamma^2]/4+i \gamma[(\omega-\omega_0(z)]} \\
   & = \frac{(\omega-\Tilde{\omega}_{R+})(\omega-\Tilde{\omega}_{R-})}{(\omega-\Tilde{\omega}_{M+})(\omega-\Tilde{\omega}_{M-})}.
\end{split}
\end{equation}
and obtain an excellent agreement for $\text{Im}[\Tilde {\omega}_R]$ in Fig.~\ref{fig:modes}(d). 

A signature of an EP is the flattening of $R(\omega)$ close to $\omega = \omega_0(z_{\text{EP}})$ \cite{Suwunnarat2021,wang2021}. Eq.~(\ref{eq:r}) demonstrates that $R(\omega)$ scales as $R(\omega) \sim [\omega-\omega_0(z_{\text{EP}})]^4/\gamma^4$ at an EP and therefore features a quartic line shape. This is confirmed in Fig.~\ref{fig:setup}(e) in reflection and Fig.~\ref{fig:modes}(a) in transmission. We emphasize that $\gamma$ decreases with increasing internal reflection so that the line shape is narrower for the sample with two dielectric slabs.

\begin{figure}
    \centering
    \includegraphics[width=8.5cm]{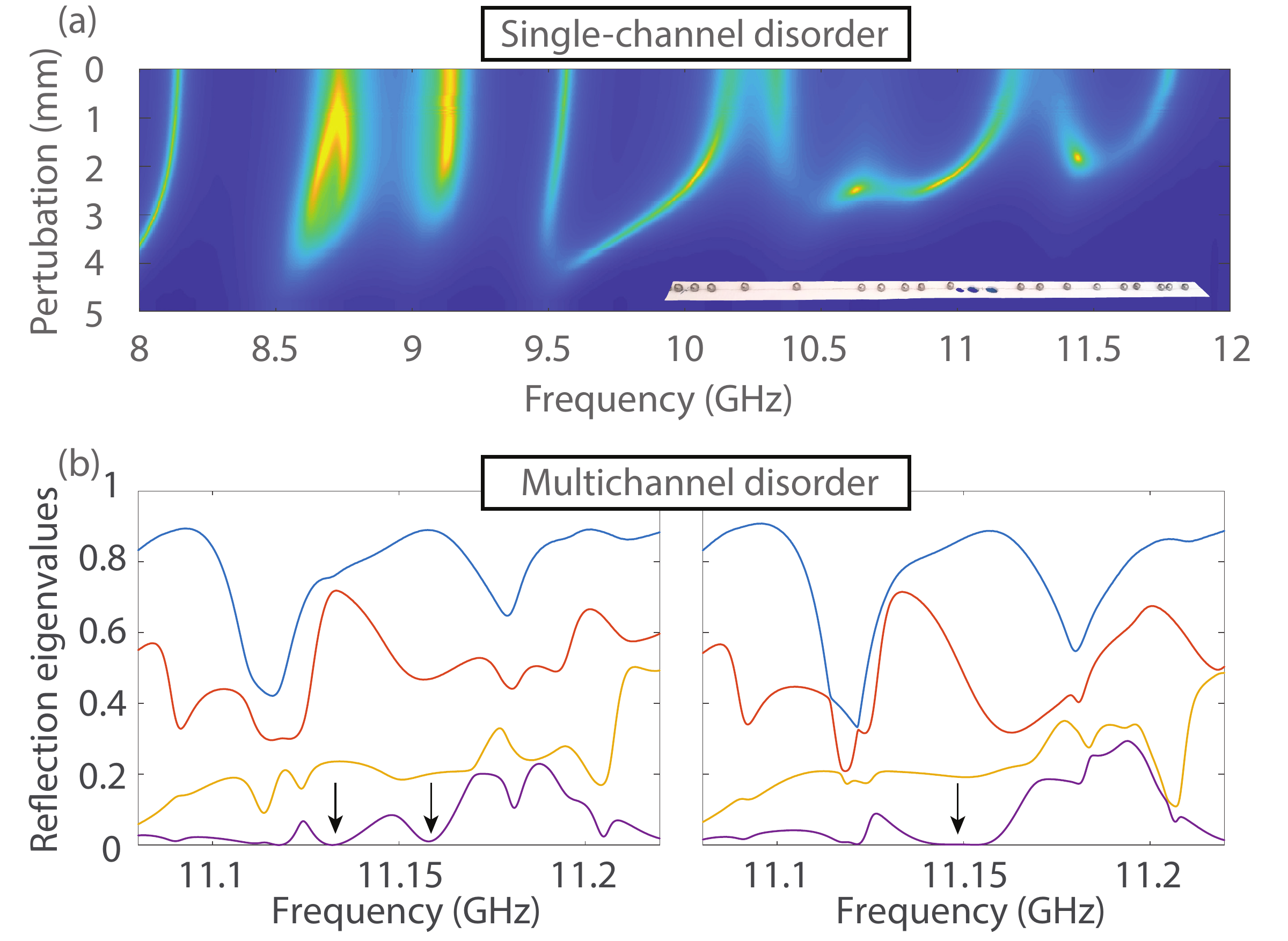}
    \caption{(a) Colorscale representation of $R(\nu,z)$ for a symmetric single-channel disordered medium. The inset shows a picture of metallic spheres inserted within the waveguide. (b) Spectrum of reflection eigenvalues in a multichannel disordered cavity in the underperturbed regime (left) and tuned at an RL-EP (right). The arrows indicate the frequencies of the coalescing zeros of $\tau_4(\nu)$. At the EP, the smallest eigenvalue displays the expected broadband quartic line shape.%(c) Variations of the spacing between two resonances (blue line) and  
    }
    \label{fig:Bscan_DisorderMedia}
\end{figure}

RL-EPs are not restricted to regular systems but exist in any complex multiple-scattering system. We now add a symmetric disorder made of 28 metallic spheres of diameter 5~mm within the empty waveguide (see Fig.~\ref{fig:Bscan_DisorderMedia}(a)). The spacing between two spheres on the left side is drawn from a uniform random distribution and we replicate this disorder with a mirror symmetry on the right side. The spectrum $R(\nu)$ still presents clear dips corresponding to RL-states but the spacing between two dips is now random. %This is reminiscent of the random distribution of the spacing between resonances in random samples. %which approaches a negative exponential (uncorrelated resonances) in localized samples. 
As the perturbation is symmetrically inserted, pairs of these dips collapse to give rise to RL-EPs. $z_\text{EP}$ is also a random variable which reflects the random field distribution within the waveguide. In contrast to the Fabry-Perot cavity case, however, not all RL-eigenvalues are on the real axis for $z=0$. An interesting case is the variation of RL-states around 8.73~GHz. Two eigenvalues first split at an EP for $z=1$~mm, move away along the real axis and then come back together to form a second EP for $z=3.8$~mm.

We also realize an RL-EP in a symmetric multichannel systems. Two arrays of $N=4$ transitions operating between 11 and 16 GHz are attached to an effectively two-dimensional rectangular cavity with metallic boundary conditions of length $L = 0.5$~m, width $W = 0.25$~m and height $h=8$~mm (see Refs.~\cite{Davy2015NCOM,Davy2021} and SM for details). Spectra of the $N \times N$ reflection matrices $r(\nu)$ are measured both at the left and right interfaces of the cavity. The cavity is made disordered with 6 aluminum rods symmetrically placed on each side of the cavity. By tuning the penetration depth of two rods in the same way as for the single channel case, we identify an RL-EP on the last eigenvalue $\tau_4(\nu,z)$ of $r^\dagger r$ at 11.15~GHz. The two zeros of $\tau_4(\nu,z)$ coalesce at the RL-EP with a characteristic quartic line shape (see Fig.~\ref{fig:Bscan_DisorderMedia}(b) and SM for the colorscale representation). We have verified that the same RL-EP is found on the reflection matrix at the other side of the cavity. 

The incident wavefront $v_R$ corresponding with an RL-state is non-trivial in complex scattering systems and can be given by the eigenvector of $r^\dagger r$ with the smallest (near-zero) eigenvalue. When the spacing $\delta$ between two zeros is large, the corresponding eigenvectors are independent wavefronts. The degree of correlation between eigenvectors $C = |v_{R+} v_{R-}^\dagger|$ is equal to 0.55 for $\delta = 0.03$~GHz. As we approach an RL-EP, the eigenvectors become strongly correlated with $C \rightarrow 1$ as the two RL-states coalesce. A single eigenvalue $\tau_4(\nu,z)$ with a flattened shape is close to zero and no decrease is observed on $\tau_3(\nu,z)$. Note that the case of diabolical point with non-degenerate states \cite{Moiseyev2011} would be dramatically different. The orthogonality of the two states within the sample would lead to two reflection eigenvalues $\tau_3$ and $\tau_4$ being simultaneously close to zero with independent eigenvectors.

 %This eigenvector is the projection of the RL-state onto the incident channels. 
%By tracking the two coalescing near-zero eigenvalues with respect to the frequency and penetration depth $z$, we now compute the correlation between eigenvectors corresponding to two RL-states $C = |v_R(\omega_{R1}) v_R^\dagger(\omega_{R2})|$ with respect to the spacing between RL-eigenvalues. When the spacing is large, the eigenvectors are independent with a minimal correlation coefficient of 0.55. However, as we approach an RL-EP, the eigenvectors become strongly correlated with $C \rightarrow 1$ as the two RL-states coalesce (see Fig.~\ref{fig:Bscan_DisorderMedia}(c)). Note that the case of diabolical point with non-degenerate states \cite{Moiseyev2011} would be dramatically different. The orthogonality of the two states within the sample would lead to two reflection eigenvalues being simultaneously close to zero with independent eigenvectors.

\begin{figure}
    \centering
    \includegraphics[width=8.5cm]{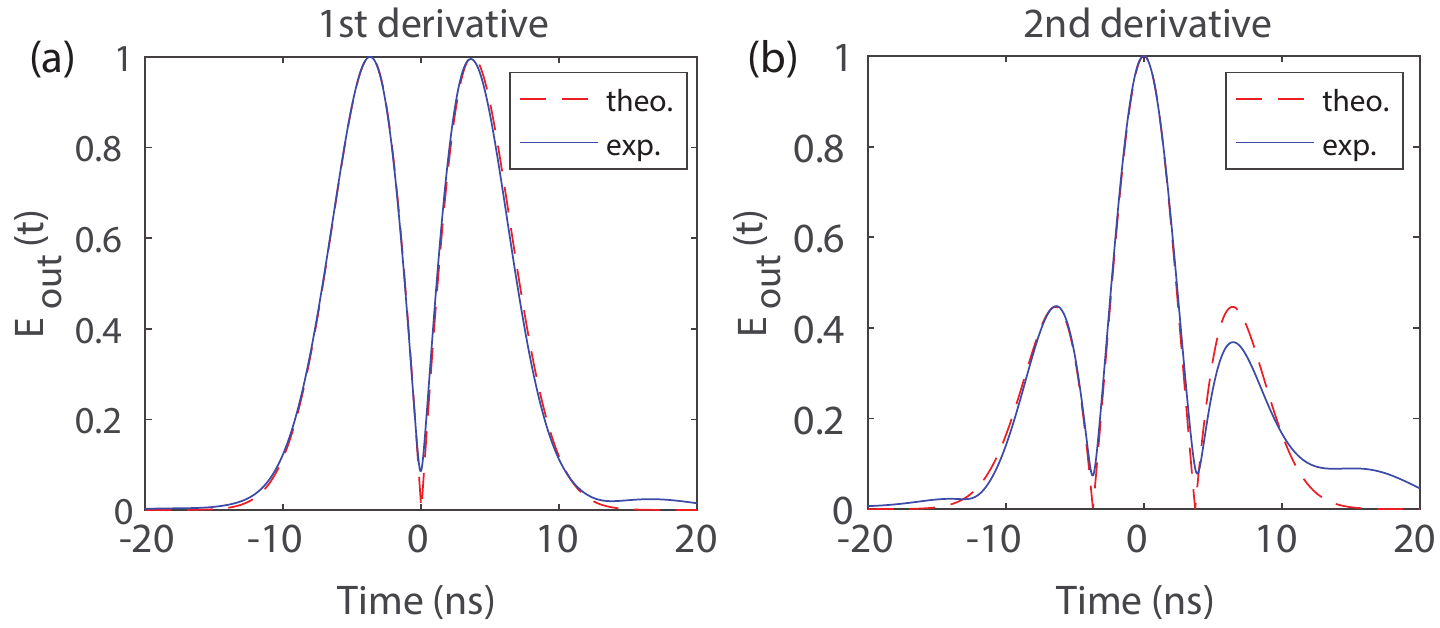}
    \caption{Analog differentiation of first (a) and second (b) order based on the transfer function in reflection associated with a RL state and a RL-EP state, respectively, of the 1D disordered waveguide. The figures display the output signal envelopes calculated for the injection of a Gaussian pulse (45~MHz bandwidth). The carrier frequencies are 8.72~GHz (a) and 10.04~GHz (b) with  $z = 5.9$~mm (a) and $z = 4.49$~mm (b), respectively.}
    \label{fig:analog}
\end{figure}

Finally, we note that the transfer functions $r(\nu)$ associated with an RL state and an RL-EP state in reflection coincides with those of a first- and second-order differentiator, respectively. Wave-based signal processing holds the promise of being fast and energy-efficient. Implementations of higher-order derivatives have been proposed mainly in carefully engineered static optical fiber systems~\cite{ngo2004new,hsue2004implementation,kulishov2005long,park2007ultrafast,li2009arbitrary}. Recently, Ref.~\cite{sol2021meta} proposed to leverage the unprecedented flexibility of complex scattering systems tuned to CPA in order to perform meta-programmable analog differentiation and implemented higher-order differentiation by cascading two systems tuned to CPA in order to emulate the transfer function of a CPA-EP. Here, by controlling the depth of the perturber penetration, we can tune our system to either an RL or RL-EP state. Thereby, we implement the second-order differentiator transfer function without using any non-linear components. For the prototypical example of a Gaussian input pulse, we calculate the envelopes of the output signal based on the measured transfer function for these two states. The results are displayed in Fig.~4 and in agreement with the analytically expected derivative. We attribute the imperfections to the fact that absorption prevents our RL-EP from lying directly on the real axis.

%Finally, we consider a direct application of our work to analog computing for wave-based differentiation. For a single zero of the reflection coefficient at $\omega_R$, the reflection coefficient scales as $r(\omega) \propto i(\omega-\omega_R)$. This transfer function satisfies the necessary condition to compute the derivative of the envelope of an incident function in the time-domain. For an incident signal $E_{\mathrm{in}}(t) = e(t) e^{i\omega_Rt}$, the reflected signal writes $E_{\mathrm{out}}(t) \propto de(t)/dt e^{i\omega_Rt}$. When the system is then tuned at an RL-EP, the scattering parameter now scales as $r(\omega) \propto -(\omega-\omega_R)^2$ which is characteristic of the second derivative of this envelope, $E_{\mathrm{out}}(t) \propto d^2e(t)/dt^2 e^{i\omega_Rt}$. The possibility to compute 1st- and 2nd-order derivatives is now demonstrated using the 1D disordered medium presented in Fig.~\ref{fig:Bscan_DisorderMedia}(a). We compute analytically the outgoing time signal for an incident Gaussian pulse with a bandwidth of 45~MHz for a single RL-state and a RL-EP. The reflected signal in the time-domain is in good agreement with the theoretical function. A slight asymmetry on the 2nd derivative is however observed [...].  

%QUANTIFY ABSORPTION, AT LEAST IN SM

In conclusion, we have observed the coalescence of RL-states into EPs in regular and disordered single- and multi-channel scattering systems. We have shown that an RL-EP in a symmetric medium requires that the spacing between two resonances is perfectly balanced by the coupling rate of symmetric and anti-symmetric modes. An interesting extension of our work would consist in observing EPs related to transmissionless states \cite{Kang2021}. These states can be found in two and three-dimensional scattering systems which theoretically provide the same behavior (broadened lineshape of the transmission spectrum and enhanced sensitivity of the zeros under perturbation of the system).

\section{Acknowledgments}
\noindent This publication was supported by the European Union through the European Regional Development Fund (ERDF), by the French region of Brittany and Rennes M{\'e}tropole through the CPER Project SOPHIE/STIC \& Ondes. M. D. acknowledges the Institut Universitaire de France. C. F. acknowledges funding from the French ``Minist{\`e}re de la D{\'e}fense, Direction G{\'e}n{\'e}rale de l'Armement''.

\section{Data availability}
\noindent The data that support the plots within this paper and other findings of this study are available from the corresponding authors on reasonable request.

\clearpage

%\section*{SUPPLEMENTAL MATERIAL}
\renewcommand{\thefigure}{S\arabic{figure}}
\renewcommand{\theequation}{S\arabic{equation}}
\setcounter{equation}{0}
\setcounter{figure}{0}
\setcounter{section}{0}

%\author{Matthieu Davy}
%\affiliation{Univ.\ Rennes, CNRS, IETR (Institut d'{\'E}lectronique et des Technologies du num{\'e}Rique), UMR–6164, F-35000 Rennes, France}
%\author{Matthias K\"uhmayer}
%\affiliation{ Institute for Theoretical Physics, Vienna University of Technology (TU Wien), A-1040 Vienna, Austria}
%\author{Sylvain Gigan}
%\affiliation{ Laboratoire Kastler Brossel, Universit{\'e} Pierre et Marie Curie, {\'E}cole Normale Sup{\'e}rieure, CNRS, Coll\`{e}ge de France, F-75005 Paris, France}
%\author{Stefan Rotter}
%\affiliation{ Institute for Theoretical Physics, Vienna University of Technology (TU Wien), A-1040 Vienna, Austria}
\date{\today}

\title{Supplementary Material for 'Exceptional points of PT-symmetric reflectionless states in complex scattering systems''}
\begin{titlepage}
\maketitle
\end{titlepage}

\section{Experimental Setups}
\subsection{Multi-Mirror Fabry-Perot Interferometer}
\begin{figure}
    \centering
    \includegraphics[width=8.5cm]{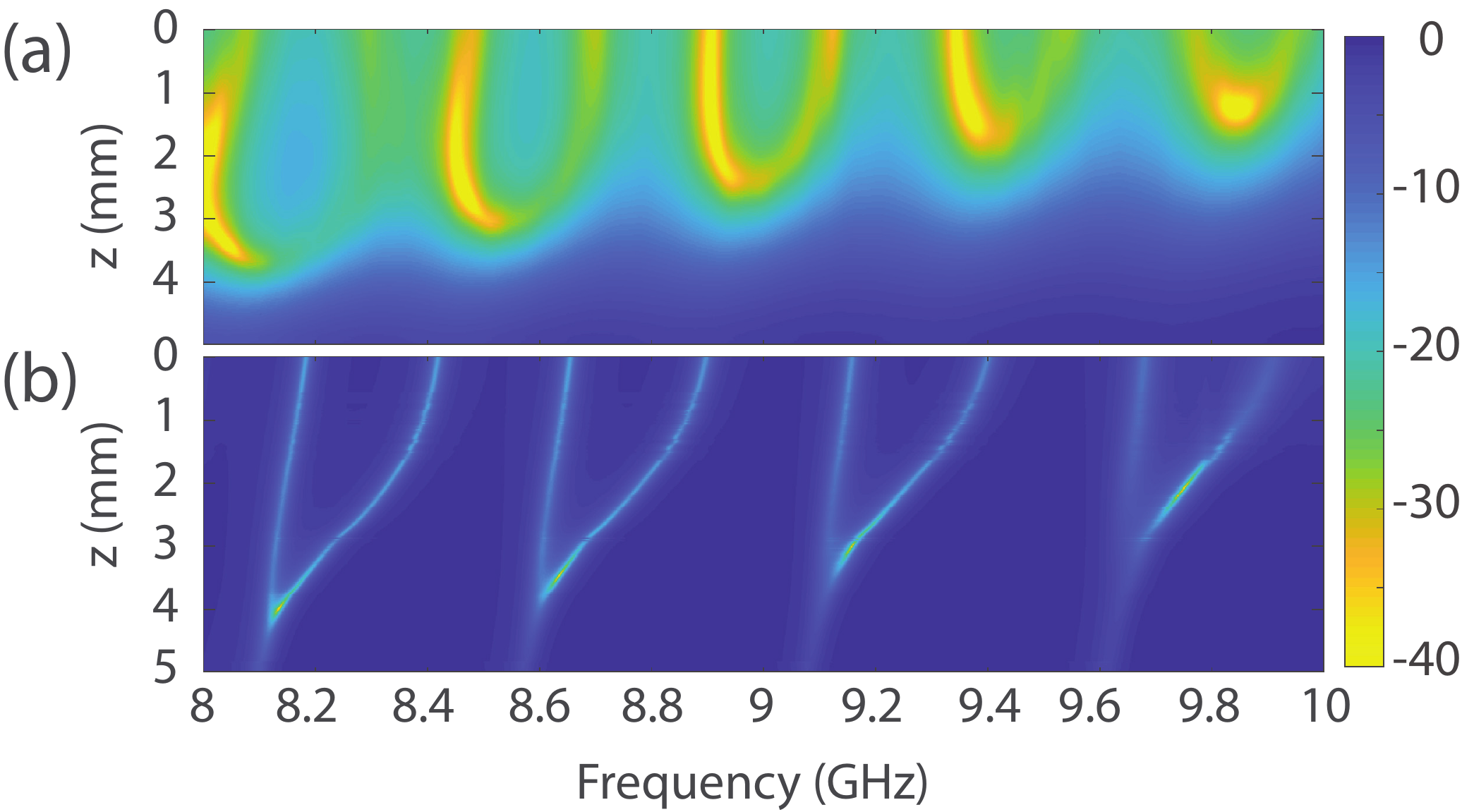}
    \caption{Colorscale representation of the reflection on a dB-scale, $10 \mathrm{log}[R(\nu,z)]$, measured with a Vector Network Analyzer between 8 and 10 GHz with zero (a) and two (b) alumina slabs at each interface.}
    \label{fig:Bscan_0Dielect_2Dielect}
\end{figure}
In Fig.~1(b,c) of the main text, we have presented a colormap of the reflection $R(\nu,z) = |r(\nu,z)|^2$ with respect to the frequency $\nu$ and the depth $z$ of the two rods inserted within the waveguide when a single dielectric alumina slab was placed between each coax-to-waveguide transition and the waveguide. The aim of this slab is to enhance the reflectivity at the interface and the Q-factor of the resonances. The collapse of RL states into RL-EPs however does not depend on this reflectivity and is a more general phenomenon. In Fig.~\ref{fig:Bscan_0Dielect_2Dielect}, we show the same colormap for 0 and 2 dielectric slabs at each interface.

Remarkably, the system still acts as a Fabry-Perot interferometer even without the slabs with regularly spaced RL-frequencies in absence of the perturbation. The reflectivity is very small but non-vanishing as the leads (coax-to-waveguide transitions) are well but not perflectly matched to the waveguide. The reflection average over the frequency range and the Q-factor of the resonances are small leading to broad zeros in reflection.  In contrast, for 2 dielectric slabs, the resonances with large Q-factors are peaked and the EPs are narrower. In both cases, As $z$ increases, RL-EPs can be oberseved as $z$ increases.

\begin{figure}
    \centering
    \includegraphics[width=8.5cm]{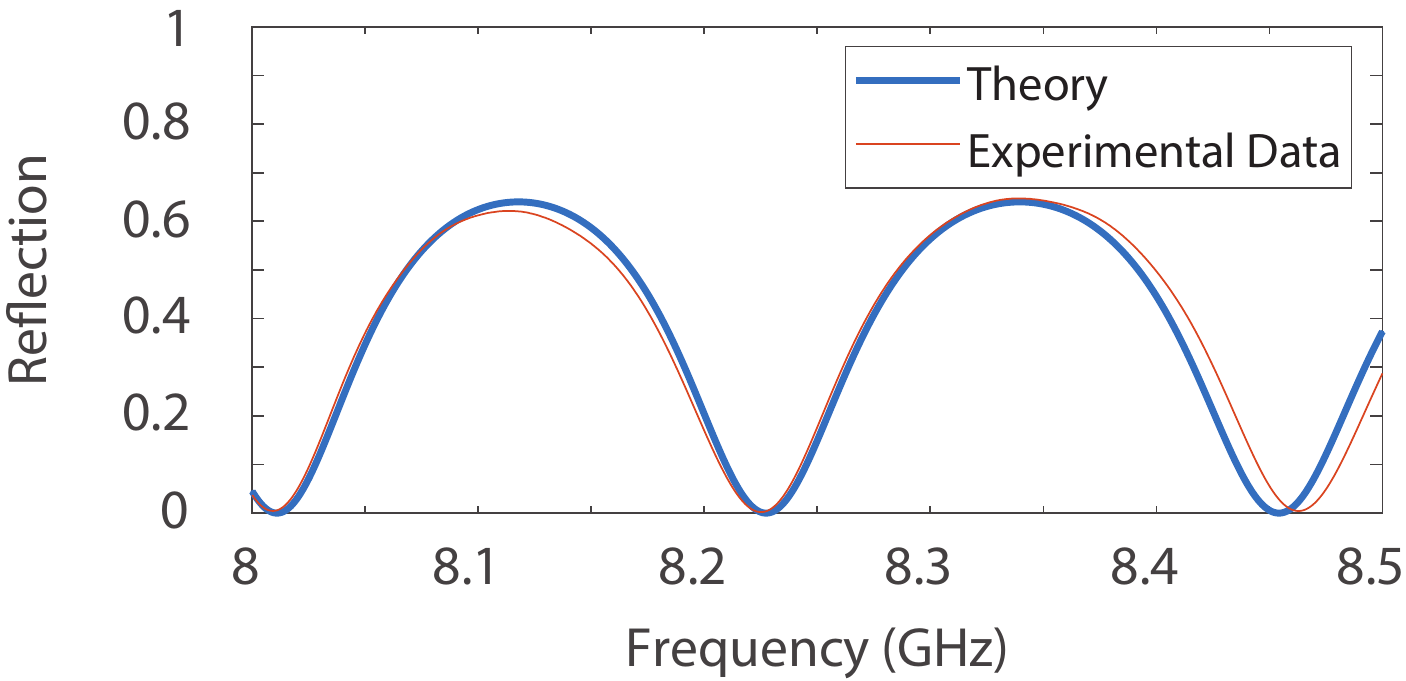}
    \caption{Spectrum of the numerical (blue) and experimental (red) reflection for an empty waveguide with one alumina slab at each interface.}
    \label{fig:R0_coupling}
\end{figure}

We estimate the reflectively $R_0$ at the interfaces provided by 0, 1 or 2 dielectric slabs by fitting the spectrum of the reflection $R(\nu,z=0)$ for an empty waveguide with its theoretical formula for a Fabry-Perot cavity

\begin{equation}
    R =1- \frac{(1 - R_0)^2}{1 + R_0^2 + 2R_0\cos{\Delta\phi}},
    \label{eq:T_FabryPerot}
\end{equation}

\noindent with $\Delta\phi = 2kL$ quantifying the phase shift of the wave propagating from one Fabry-Perot mirror to the other. Here $k=\frac{2\pi}{\lambda}$ is the wave number and $L$ the distance between the mirrors. The spectrum of the reflection and its best fit for a single dielectric slab placed at each interface is shown in Fig.~\ref{fig:R0_coupling}. We estimate that $R_0 = 0.002$, $R_0 = 0.25$ and $R_0 = 0.46$ for 0, 1 and 2 slabs, respectively.

\subsection{Disordered one-dimensional medium}
\begin{figure}
    \centering
    \includegraphics[width=8.5cm]{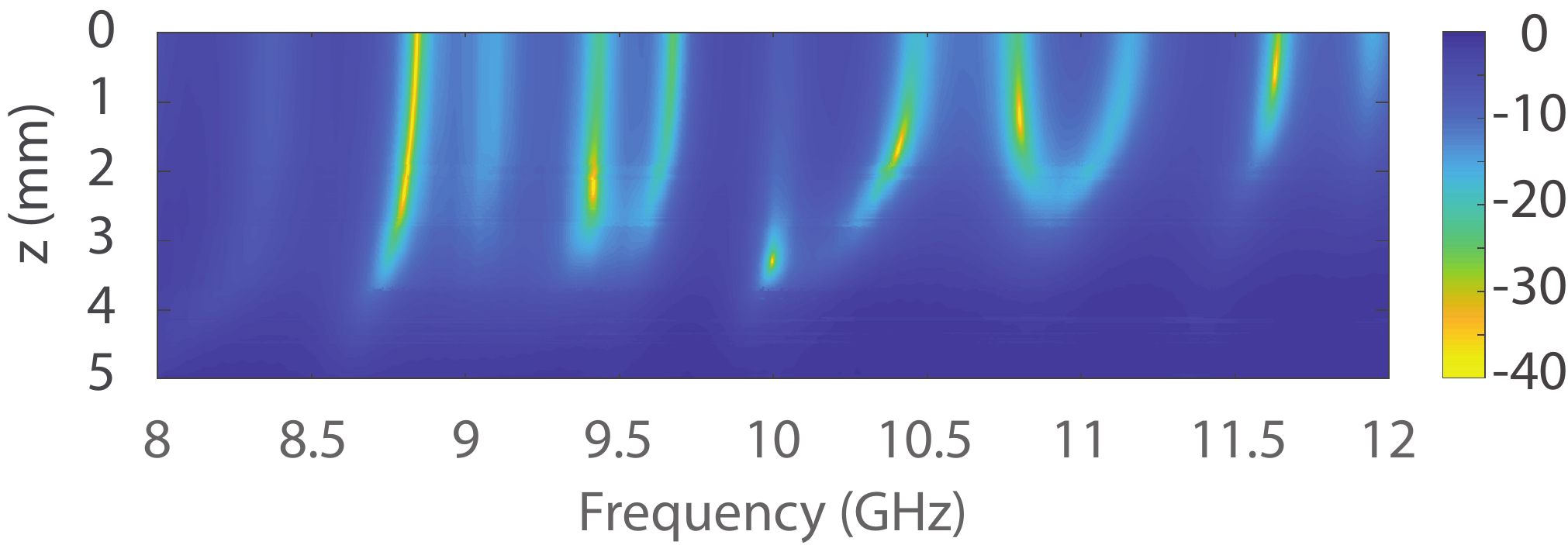}
    \caption{Colorscale representation of $R(\nu,z)$ for a random disordered medium without mirror symmetry.}
    \label{fig:SM_Bscan_randomDisorder}
\end{figure}

As mentioned in the main text, RL-EPs can exist in any complex multiple-scattering systems. We add inside the waveguide a random single-channel disorder made of the same 28 metallic spheres as those presented in the main text. The reflection $R(\nu,z)$ is shown in Fig.~\ref{fig:SM_Bscan_randomDisorder}. Without perturbation ($z=0$), the spectrum presents randomly spaced dips. In contrast to the Fig.~3(a), one can see less RL-EPs with the random disorder than the symmetric one. Indeed, with a symmetrical disorder, the RL-eigenvalues can leave the real axis only by coalescing with another RL-eigenvalue whereas with a random disorder, a RL-eigenvalue can randomly leave the real axis by itself.

\section{Theoretical analysis}

In this section, we provide more details on the calculations using the effective two-level Hamiltonian approach. The effective Hamiltonian writes
\begin{equation}
    H_{\text{eff}} = H_0 -\frac{i}{2} V_0 V_0^T -\frac{i}{2}V_1 V_1^T,
\label{eq:Heff}
\end{equation}
where the left and right vectors $V_0$ and $V_1$ account for the coupling of the channels to the closed system. In the basis of two modes of the closed waveguide, $H_0$ is a diagonal matrix with elements $\omega_0 -\delta\omega_0/2$ and $\omega_0 +\delta\omega_0/2$. We have identified that the coupling vectors are symmetric, $V_0 = \sqrt{\gamma/2}(1~1)$ and anti-symmetric $V_1 = \sqrt{\gamma/2}(1~-1)$, respectively (see main text). $\gamma$ is the coupling rate of the channels to the cavity and therefore on the reflectivity at the interfaces of the waveguide. This yields
\begin{equation}
    H_{\text{eff}} = \omega_0 \id + \frac{1}{2}\begin{pmatrix} - \delta \omega_0 -i\gamma   & 0\\ 0  & \delta \omega_0 - i\gamma  \end{pmatrix} 
\label{eq:Heff}
\end{equation}
Its eigenvalues are then
\begin{equation}
    \Tilde {\omega}_{M\pm} = \omega_0 \pm \delta\omega_0/2 - i\gamma/2
\label{eq:eigenvalues_modes}
\end{equation}

The scattering matrix $S(\omega)$ can then be derived from the effective Hamiltonian using $S(\omega) = 1-iV^T [\omega \id - H_{\text{eff}}]^{-1} V$, where $V = [V_0 V_1]$. The reflection coefficient $r(\omega)$ at the left interface is especially given by $r = 1-iV_0^T [\omega \id - H_{\text{eff}}]^{-1} V_0$. Using that 
\begin{equation}
     [\omega \id - H_{\text{eff}}]^{-1} = \begin{pmatrix} \frac{1}{\omega-\omega_0+\delta\omega_0/2+i\gamma/2} & 0\\ 0  &  \frac{1}{\omega-\omega_0-\delta\omega_0/2+i\gamma/2}  \end{pmatrix} 
\label{eq:Heff}
\end{equation}
we find
\begin{equation}
   r(\omega) = 1-\frac{i\gamma/2}{\omega-\omega_0+\frac{1}{2}[\delta\omega_0+i\gamma]}-\frac{i\gamma/2}{\omega-\omega_0+\frac{1}{2}[-\delta\omega_0+i\gamma]}.
   \label{eq:r}
\end{equation}
This equation can then be simplified to
\begin{equation}    \label{eq:r}
   r(\omega) = \frac{[\omega-\omega_0]^2+[\gamma^2-\delta\omega_0^2]/4}{[\omega-\omega_0]^2-[\delta\omega_0^2+\gamma^2]/4+i \gamma[(\omega-\omega_0]}.
\end{equation}
which is Eq.~(3) of the main text.

Simple calculations show that the expression of $r(\omega)$ can also be factorized in terms of eigenfrequencies $\Tilde {\omega}_{M\pm}$ and RL-eigenvalues $\Tilde {\omega}_{R\pm} = \omega_0(z) \pm \frac{1}{2}\sqrt{\delta\omega_0(z)^2 -\gamma^2}$ as 
\begin{equation}
   r(\omega)  = \frac{(\omega-\Tilde{\omega}_{R+})(\omega-\Tilde{\omega}_{R-})}{(\omega-\Tilde{\omega}_{M+})(\omega-\Tilde{\omega}_{M-})}.
   \label{eq:r}
\end{equation}
As expected, $r(\omega)$ vanishes when $\omega = \Tilde{\omega}_{R\pm}$. 

\section{Modal Analysis}
\subsection{Harmonic Inversion}
To extract the resonances $\Tilde{\omega}_n = \omega_n -i\Gamma_n/2$ of the system, we perform a modal analysis between 7 and 11 GHz of the transmission coefficient. We seek to decompose $t(\omega)$ as 
\begin{equation}
   t(\omega)  = \Sigma_n \frac{t_n}{\omega-\omega_n+i\Gamma_n/2}.
   \label{eq:r}
\end{equation}
The coefficients $t_n$ are the modal transmission coefficients that are associated with each resonance. We use the Harmonic Inversion (HI) method applied to the inverse Fourier transform of $t(\omega)$ in the time domain \cite{Mandelshtam1997,Davy2018}. An excellent agreement with experimental results is found with 19 modes.

\begin{figure}
    \centering
    \includegraphics[width=8.5cm]{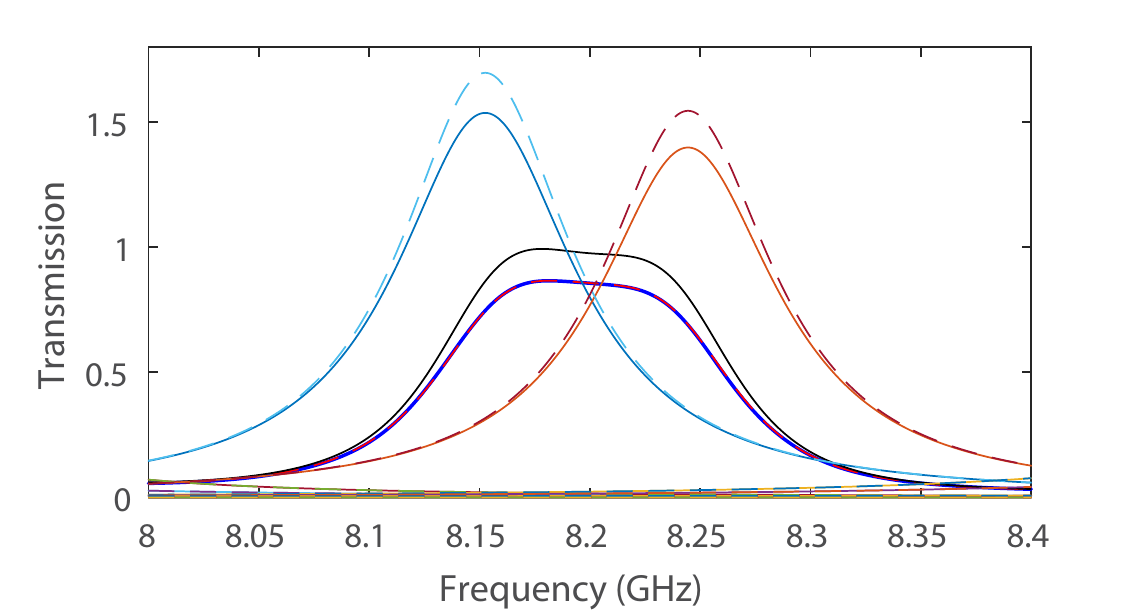}
    \caption{The transmission spectrum (blue line) and its reconstruction using the modal analysis (red dashed-line). The corresponding modes are shown in blue and orange. By compensating the absorption strength, $\Gamma_n \rightarrow \Gamma_n - \Gamma_a$ with $\Gamma_a = 30$~MHz, the maximum of the reconstructed transmission (black line) reaches unity. The dashed blue and orange lines are the contributions of the two modes when the absorption is removed.  }
    \label{fig:absorption}
\end{figure}

Uniform losses within the sample broaden the linewidth, $\Gamma_n \rightarrow \Gamma_n + \Gamma_a$, and reduces transmission through the sample. We now estimate the absorption rate $\Gamma_a$ at a RL-EP in the case of a perturbed Fabry-Perot interferometer with one dielectic slab placed at each interface. The maximum transmission found experimentally at the RL-EP is 0.865. We compensate the linewidth in the reconstructed spectrum of $T(\omega))$ and find that the transmission reaches unity for $\Gamma_a = 30$~MHz, as shown in Fig.\ref{fig:absorption}.

We observe that a slight asymmetry on the flattened line shape of the transmission spectrum found experimentally at the EP. The strengths of the two modes contributing to the transmission are indeed not equal. Their maxima are $T_n = |t_n|^2= 1.536$ and $T_{n+1}=1.398$. We attribute this difference to the joint effects of a non-perfect symmetry of the system due to inevitable fabrication errors and a non-uniform absorption strength. Note that the modal strength exceeds unity as a consequence of the bi-orthogonality of quasi-normal modes in open systems \cite{Davy2019PRR}. In symmetric systems, $T_n$ is equal to the Petermann factor of the modes, $T_n = K_n$ which is a measure of the degree of complexness of the eigenfunctions \cite{Poli2010}. For isolated resonances, $K_n ~sim 1$ but $K_n$ increases with the modal overlap between resonances. Here, two resonances are overlapping yielding $T_n = K_n >1$.

\subsection{Spacing between resonant frequencies}

\begin{figure}
    \centering
    \includegraphics[width=8.5cm]{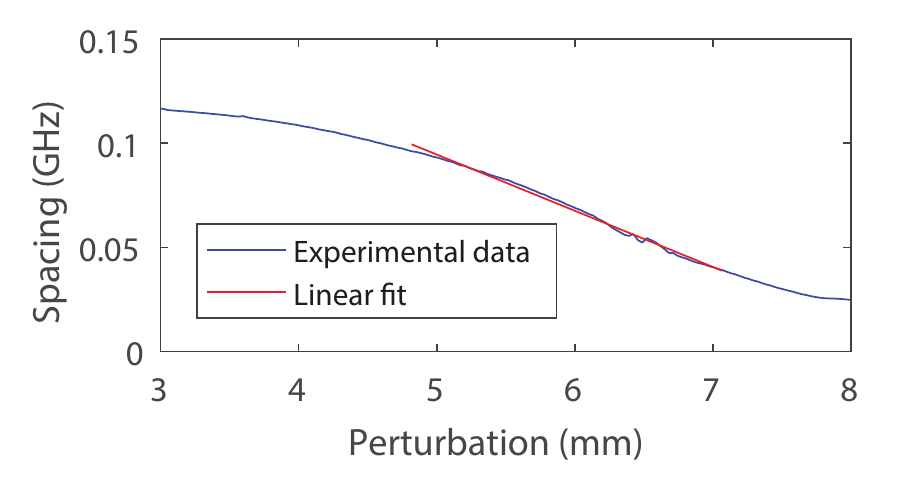}
    \caption{Spacing $\delta\omega_0(z)$ between central frequencies of the two resonances close to 8.2~GHz for a multimirror cavity with one dielectric slab on each interface. Near the EP at $z_{\mathrm{EP}}=6.2$~mm, the experimental data (blue line) are in good agreement with a linear fit (red line) as a function of $z$.}
    \label{fig:spacing}
\end{figure}

In Fig.~\ref{fig:spacing}, the spacing between two resonant frequencies $\delta\omega_0(z)$  is shown as a function of the penetration depth $z$ for a Fabry-Perot cavity with a a single alumina slab placed at each interface. Near the EP at $z_{\mathrm{EP}} = 6.8$~mm, $\delta\omega_0(z)$ is seen to be well approximated with a linear function. Using that $\delta\omega_0(z_{\mathrm{EP}}) = \gamma$, we can write $\delta\omega_0(z) = \gamma + \kappa/2 (z_{\mathrm{EP}}-z)$, where $\kappa$ is a positive coefficient. The spacing between real RL-eigenvalues $\delta \Tilde {\omega}_R = \sqrt{\delta\omega_0(z)^2 -\gamma^2}$ gives $\delta \Tilde {\omega}_R = \sqrt{\kappa\gamma (z_{\mathrm{EP}}-z) + \kappa^2/4 (z_{\mathrm{EP}}-z)^2}$. In the vicinity of the EP, the square root behavior $\delta \Tilde {\omega}_R \sim \sqrt{\kappa\gamma (z_{\mathrm{EP}}-z)}$ reflects the square-root sensitivity of EPs to an external perturbation of the system. The same is obtained for $z>z_{\mathrm{EP}}$ on the splitting of the imaginary parts of the two complex-conjugate RL-eigenvalues.

% \begin{figure}
%     \centering
%     \includegraphics[width=8.5cm]{Figures/FIG_RESONANCES_2.pdf}
%     \caption{Resonances of the reflection $R(\nu,z)$ measured between 8.5 and 8.8~GHz for different air gap between the top of the waveguide and the third rod. }
%     \label{fig:EPshifted}
% \end{figure}
% This is confirmed in Fig.~\ref{fig:EPshifted} which represents the resonances as a function of the initial perturbation for three configurations corresponding to different air gap, $\Delta_d$, between the top of the waveguide and the third rod.

\section{Multichannel cavity}
\begin{figure}
    \centering
    \includegraphics[width=8.5cm]{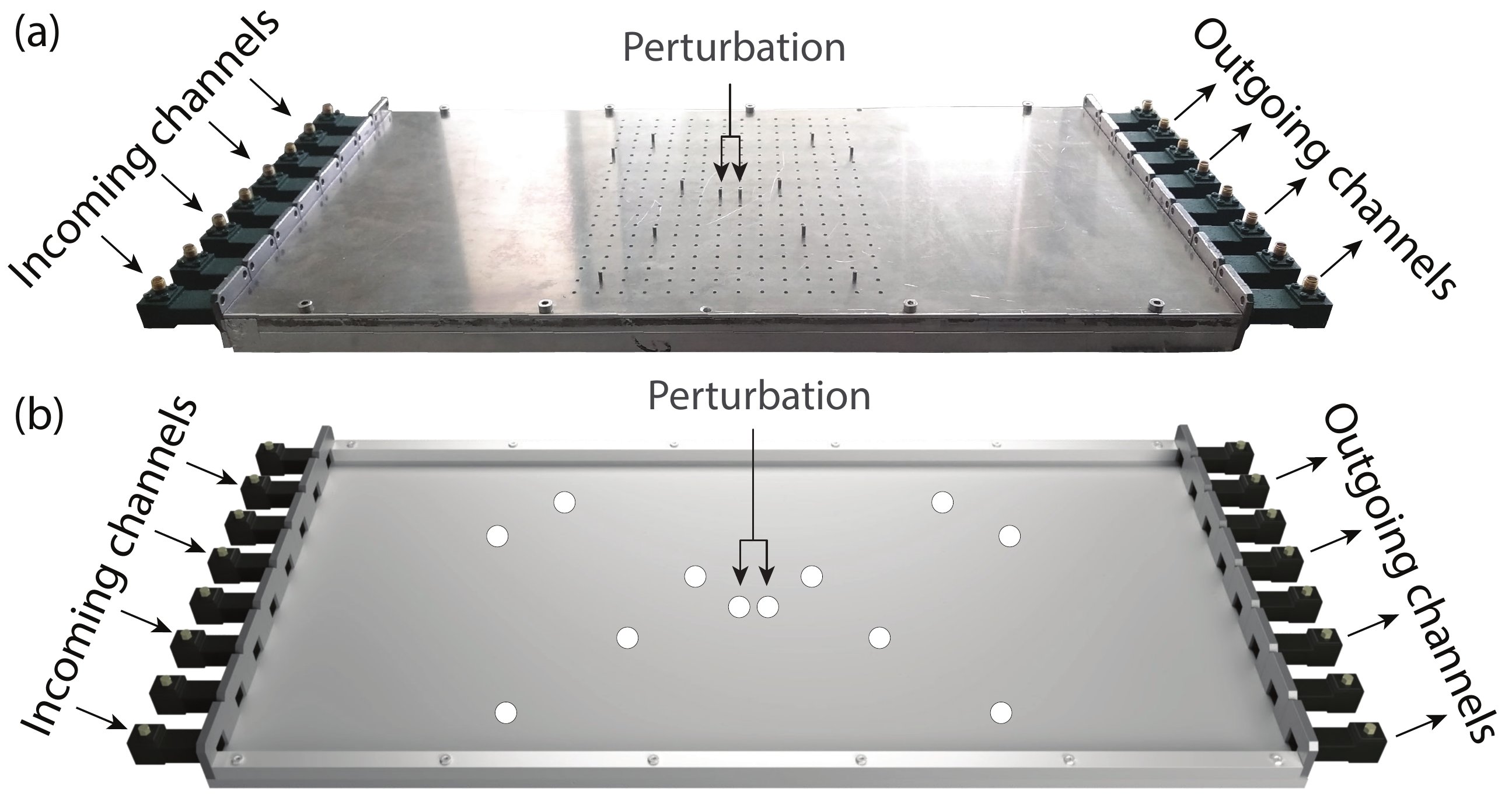}
    \caption{(a) Photography of the Multi-channel cavity. (b) Sketch of the cavity for which the top has been removed to see the symmetric disorder inside the cavity.}
    \label{fig:MultiChannelCavity}
\end{figure}
\begin{figure}
\centering
\includegraphics[width=8.5cm]{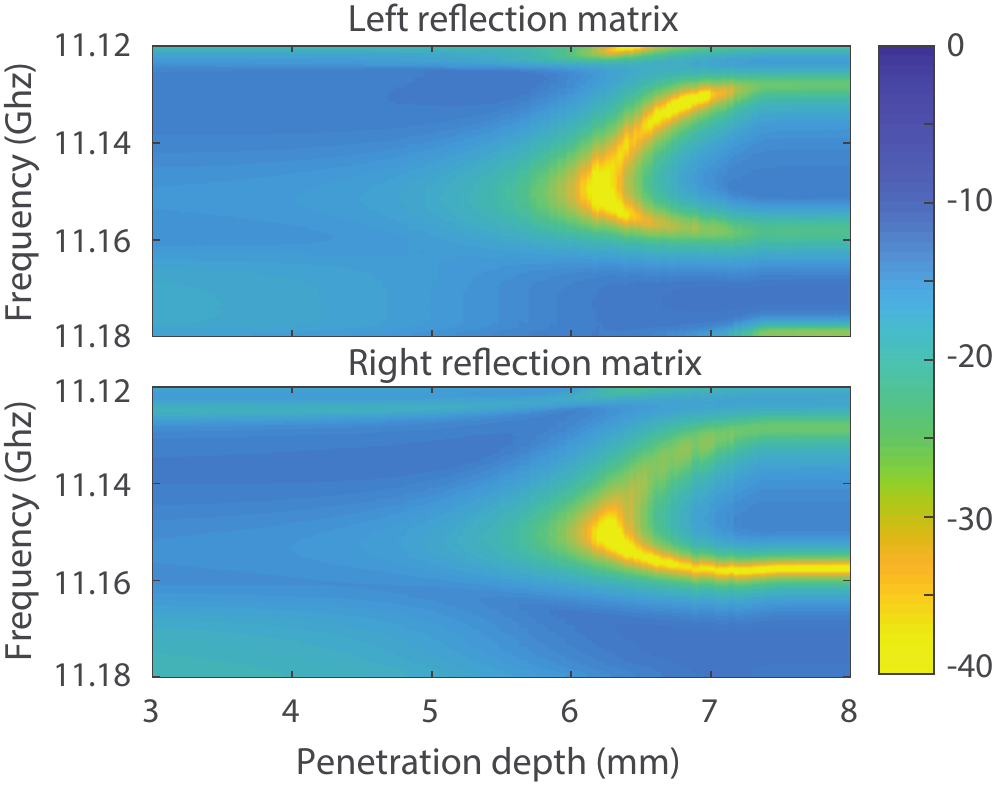}
\caption{Colorscale representation of the last reflection eigenvalue $\tau_4(\nu,z)$ on a dB-scale for the multichannel case. The minimal reflection is shown for the left (top) and right (bottom) reflection matrices.}     
\label{fig:symm} 
\end{figure}
The reflection coefficient $r(\nu)$ and the transmission coefficient $t(\nu)$ are measured through a rectangular multi-channel cavity (length $L=500$~mm, width $W=250$~mm and height $H=8$~mm) between two arrays of $N=4$ single-channel. Note that $8$ transitions are observed at each side of the cavity on Fig.~\ref{fig:MultiChannelCavity}. However, only 4 of them are used to measure the reflection matrix. The others are not connected to the VNA. This open-circuit condition is similar to a metallic boundary conditions for the cavity.

An array of $14 \times 20$ holes are drilled symmetrically with respect to the center on the top plate of the cavity with a spacing of $12$~mm between each hole (see Fig.\ref{fig:MultiChannelCavity}(a)). Ten aluminium rods are inserted symmetrically through these holes thus creating a symmetric disorder inside the cavity (see Fig.~\ref{fig:MultiChannelCavity}(b)). As with the single mode waveguide presented in the main text, a perturbation is then introduced in the cavity by inserting symmetrically two aluminium rods spaced by $12$~mm near to the center. The penetration depth of these two rods is tuned to observe the RL-EPs.
 
Due to its large dimensions and inevitable fabrication errors, the cavity is not perfectly symmetric. However, we verify that the RL-EP presented in the main text is found on the both the left and right reflection matrices. In Fig.~\ref{fig:symm}, we present the colorscale representation of the smallest reflection eigenvalues $\tau_4(\nu,z)$ for both matrices. The multichannel RL-EP can be clearly observed on the two figures as expected.

%\nocite{*}

\bibliographystyle{apsrev4-1}
\providecommand{\noopsort}[1]{}\providecommand{\singleletter}[1]{#1}%

\end{document}